\documentclass[prb,twocolumn]{revtex4}

\usepackage{epsfig}
\usepackage{amsmath}

\begin{document}
\title{Dynamics and (de)localization in a one-dimensional tight-binding
chain}
\author{Antonio \v{S}iber}
\email{asiber@ifs.hr}
\affiliation{Institute of Physics, P.\ O.\ Box 304, 10001 Zagreb, Croatia}

\begin{abstract}
A simple tight-binding model is used to illustrate how the time dependence
of a state vector can be obtained from all the eigenvalues
and eigenvectors of the Hamiltonian. The behavior of the 
eigenvalues and eigenvectors is studied for various parameters and allows
us to study scattering-like events, impurity states, and localization in
disordered systems. 
\end{abstract}

\begin{center}
Published in Am. J. Phys. {\bf 74}, 692 (2006).
\end{center}

\maketitle

\section{Introduction}
\label{sec:intro0}
For most physics students the Schr\"{o}dinger representation of quantum
mechanics is more appealing than the matrix or Heisenberg representation.
The reason is due in part to the fact that the Schr\"{o}dinger
representation is more suited for visualization.\cite{goldb,Schiff} Also
most students are more accustomed to thinking in terms of functions rather
than in terms of abstract eigenvectors.

In this article I discuss a simple but versatile tight-binding Hamiltonian
whose eigenvalues, eigenvectors and dynamics can be obtained and easily
visualized in the matrix representation. The solution of the physical system
represented by the Hamiltonian is obtained numerically by using a short
program that is given in Appendix~A. The code allows for visualization of
the spectrum of eigenvalues and eigenvectors and also the dynamics
generated by the Hamiltonian.

The model Hamiltonian of a one-dimensional chain with nearest neighbor
couplings is discussed in Sec.~\ref{sec:intro} and a complete set of states
describing the single particle dynamics is introduced.
Section~\ref{sec:phys} briefly describes how the Hamiltonian generates the
time dependence of an initial state vector and demonstrates that the
eigenvalues and eigenvectors define the dynamical behavior of the system. In
Sec.~\ref{sec:appli} I discuss several applications that are of
pedagogical interest. In particular, the dynamics of a state initially
localized on a particular site is studied. The dynamics is correlated with
the spectrum of the Hamiltonian and properties of its eigenvectors
(localized versus delocalized). Several aspects of the problem that are of 
research interest are discussed, including the problems of localization and
conductance in chain-like molecules, such as DNA. 

\section{Definition of the problem}
\label{sec:intro}

Consider the Hamiltonian:
\begin{equation}
H = \sum_i E_i c_i^{\dagger} c_i + \sum_i T_i [ c_{i+1}^{\dagger} c_i + 
c_i^{\dagger} c_{i+1} ],
\label{eq:ham}
\end{equation}
where $c_i^{\dagger}$ and $c_i$ are the creation and 
destruction operators of a particle on site $i$, respectively.
The Hamiltonian represents a chain of sites denoted by indices $i$; a
particle can hop from one site to another due to the nonvanishing values of
$T_i$ which are often called hopping matrix elements. 

We restrict our attention to a single particle (or, more generally, an
excitation) propagating through the chain. The dynamics 
can be described in a position-occupation basis, that a 
basis of states denoted by $\left \{ |i \rangle , i=1,\ldots,N \right \}$
such that
\begin{equation}
|i \rangle = |0 0 \ldots 1_i \ldots 0 \rangle,
\label{eq:pobasis1}
\end{equation}
where $i$ denotes a particular site occupied by an excitation and $N$ denotes 
the total number of sites. The action of $c$-operators in this basis is
simple:
\begin{subequations}
\label{eq:pobasis2}
\begin{align}
c_i^{\dagger} |0 \rangle &= |i \rangle \\
c_i |i \rangle &= |0 \rangle \\
c_i |j \rangle &= 0, \quad i \ne j ,
\end{align}
\end{subequations}
where we consider only singly occupied states; the vacuum is denoted by $|
0 \rangle$. In the basis that is restricted to singly occupied states, we
can equivalently (and without reference to creation/destruction operators)
represent the Hamiltonian as
\begin{equation}
H = \sum_i E_i |i \rangle \langle i | + 
\sum_i T_i \big[ |i+1 \rangle \langle i| 
+ | i \rangle \langle i+1| \big].
\end{equation}

The position-occupation basis does not diagonalize 
the Hamiltonian in Eq.~(\ref{eq:ham}), except in the trivial 
case of $T_i = 0$ for all $i$. However, it is the basis that is simplest 
conceptually and most easy to visualize.

Equation~\eqref{eq:ham} is a simplified tight-binding 
Hamiltonian and is discussed in many textbooks on condensed matter physics 
(see for example, Ref.~\onlinecite{Ashk}). Its matrix representation 
is easy to construct. The simplest and numerically most feasible 
way is to consider a tridiagonal matrix in the 
position-occupation basis (Eqs.~(\ref{eq:pobasis1}) and 
(\ref{eq:pobasis2})) as
\begin{equation}
H = 
\begin{pmatrix}
E_1 & T_1 & 0 & 0 & \cdots & 0 & 0 \cr
T_1 & E_2 & T_2 & 0 & \cdots & 0 & 0\cr
0 & T_2 & E_3 & T_3 & \cdots & 0 & 0\cr
\vdots & \vdots & \vdots & \vdots & \ddots & \vdots & \vdots \cr
0 & 0 & 0 & 0 & \cdots & E_{N-1} & T_{N-1} \cr
0 & 0 & 0 & 0 & \cdots & T_{N-1} & E_N 
\end{pmatrix} .
\label{eq:matrica}
\end{equation}
The Hamiltonian is represented in a basis of singly occupied states;
that is, we consider only this subset of states of Fock space (which
includes multiply occupied states; for example, we might want to consider
the dynamics of two excitations). Note that the
Hamiltonian defined in Eq.~(\ref{eq:ham}) cannot induce transitions between
the Fock subspaces corresponding to a different total number of excitations.

Periodic boundary conditions are not imposed in the
matrix representation in Eq.~(\ref{eq:matrica}). Periodic boundary
conditions would require nonvanishing $(1,N)$ and $(N,1)$ elements,
that is, the upper right and the lower left corner of the matrix. The
matrix in Eq.~(\ref{eq:matrica}) can be easily set up and diagonalized
numerically. It is enough to specify only two arrays of real numbers, one
of length
$N$ which contains the diagonal values of the matrix, and the
other of length $N-1$ which contains the elements of the Hamiltonian matrix
along its first subdiagonal.

\section{Time dependence of state vectors}
\label{sec:phys}

Let us assume that all the eigenvalues $\epsilon _k$ and 
eigenvectors $| \xi_k \rangle $ 
of the Hamiltonian $H$ are known 
($H | \xi_k \rangle = \epsilon _k | \xi_k \rangle$). 
We further assume
that the system is at time $t=0$, in some known or prepared
state $ |\Psi (t=0) \rangle$. The state $|\Psi (t=0) \rangle$
can be projected onto the basis of eigenvectors of the full
Hamiltonian:
\begin{equation}
|\Psi (t=0) \rangle = \sum _ k a_k |\xi_k \rangle,
\label{eq:psi0}
\end{equation}
where the projection coefficients $a_k$ are given by
\begin{equation}
a_k = \hspace{1pt} \langle \xi_k| \Psi (t=0) \rangle,
\end{equation}
because the $| \xi_k \rangle$ are assumed to be orthonormal, that is,
\begin{equation}
\hspace{1pt} \langle \xi_m | \xi_n \rangle = \delta _{m,n}.
\end{equation}
Let us denote the time evolution operator by $U(t)$;
that is, $U(t)$ acts on an arbitrary state $| \Psi (t=0)
\rangle$ and evolves it to the state $| \Psi (t) \rangle$, 
\begin{equation}
U(t) | \Psi (t=0) \rangle = | \Psi (t) \rangle.
\end{equation}
The action of the time evolution operator on the eigenvectors of the
problem is trivial\cite{Schiff2}:
\begin{equation}
U(t) |\xi_k \rangle = \exp (-i \epsilon _k t/\hbar) |\xi_k \rangle,
\end{equation}
which implies that
\begin{eqnarray}
| \Psi (t) \rangle &=& \sum _ k a_k \exp (-i \epsilon _k t/\hbar) | \xi_k
\rangle \\
&=& \sum _ k \langle \xi_k | \Psi (t=0)
\rangle 
\exp (-i \epsilon _k t/\hbar) | \xi_k \rangle .
\label{eq:tdep1}
\end{eqnarray}

We assume that the wave function (or state 
vector) is initially chosen to be localized on a particular site $l$ of the
chain, that is, $|\Psi (t=0) \rangle = |l \rangle$. We ask about the
probability that after some time $t$ the 
excitation is on some other site $m$. To obtain this information, the state
vector 
$|\Psi (t)\rangle$ must be projected onto the position-occupation basis,
that is, we should calculate $N$ projections $b_{m,l} (t)$
\begin{equation}
b_{m,l} (t) = \langle m | \Psi (t) \rangle 
= \sum _ k \hspace{1pt} \langle \xi_k | l \rangle \hspace{1mm} 
\langle m | \xi_k \rangle 
\exp (-i \epsilon _k t/\hbar).
\label{eq:tdep2}
\end{equation}
The quantity $|b_{m,l} (t)|^2$ is the probability that at time $t$, the
excitation initially localized on site $l$ is found on site $m$. A
pedagogical account of the definition of probability current in
tight-binding problems can be found in Ref.~\onlinecite{Erasmo}.

\section{Applications of the model}
\label{sec:appli}

As mentioned, the matrix in Eq.~(\ref{eq:matrica}) can be diagonalized
numerically. One possible way of doing so is described in Appendix~A, which
lists a sample program to set up the Hamiltonian matrix and diagonalize it.
The result of this numerical procedure is an array of
$N$ Hamiltonian eigenvalues $\epsilon_k$ and a $N \times N$ matrix 
(or 2D array) of eigenvectors $| \xi_k \rangle$, which provides a
complete solution of the problem, including its time dependence. In the
following I present several applications of the code.

\subsection{Regular chain, propagation of the initially localized state}
\label{regular}

\begin{figure}[h]
\centerline{
\epsfig {file=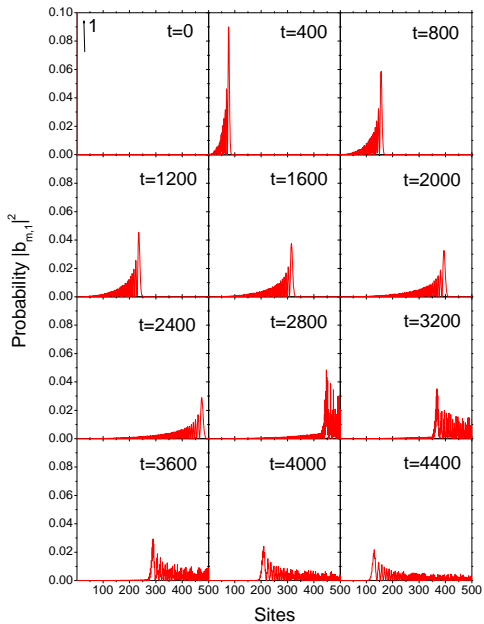,width=7cm}
}
\caption{Time dependence of the probability distribution $|b_{m,1}|^2$ 
for a completely regular chain. The excitation was initially 
localized on the first site, 
that is, $|\Psi (t=0) \rangle = |1 \rangle$. The time is measured in
units of $\hbar/e$ and the number of sites $N=500$.}
\label{fig:slika1}
\end{figure}

\begin{figure}[h]
\centerline{
\epsfig {file=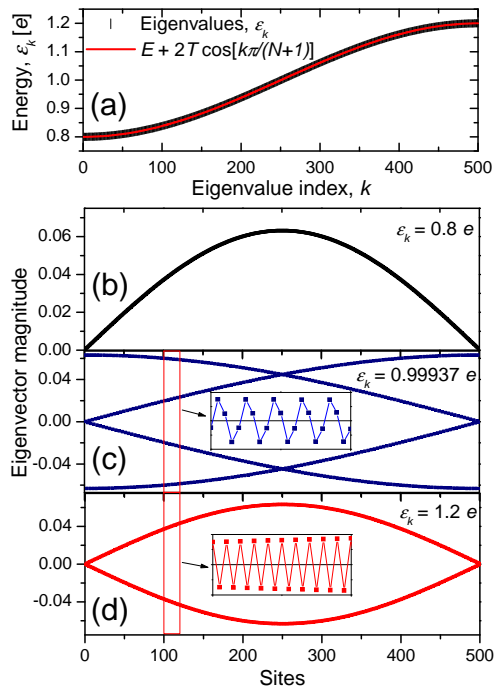,width=7cm}
}
\caption{Eigenvalues and three characteristic eigenvectors of a 
regular chain. (a) Hamiltonian eigenvalues, (b) eigenvector of 
the lowest energy eigenstate ($\epsilon_k = 0.8 e$), (c) 
eigenvector of eigenstate with energy $\epsilon_k = 0.99937 e$ (250th 
eigenstate), and (d) eigenvector of eigenstate with energy 
$\epsilon_k = 1.2 e$ (highest energy eigenstate). The insets in the 
(c) and (d) show the eigenvector magnitude between the 100th 
and 120th sites (these are blowups of the eigenvector magnitude in the 
regions bounded by the two thin rectangles in the main plots).}
\label{fig:slika1a}
\end{figure}

I first consider a completely regular chain, that is, a chain in which 
$E_i = E$ and $T_i = T$ for all $i$.  The characteristic energies
($E$) and the hopping matrix elements ($T$) are set to $E=e$ and 
$T=-0.1e$, where $e$ denotes the energy scale. Note that the chosen 
energy scale also fixes the characteristic time scale, which is 
given by $\hbar/e$ (see Eq.~(\ref{eq:tdep2})). The results of this 
calculation are given in Figs.~\ref{fig:slika1} and
\ref{fig:slika1a}. It 
is clearly observed how the initially localized state delocalizes over many
sites and eventually hits the right end of the chain, reflecting from
it. The horizontal axis in the plots is the site index which is in
principle unrelated to any characteristic length. The spatial dimensions 
are hidden in the hopping (or overlap) matrix elements $T_i$; the more 
separated the atomic orbitals $E_i$ are, the smaller their overlap and 
the corresponding hopping matrix element.\cite{Ashk}

Figure \ref{fig:slika1a} displays the eigenvalues of the Hamiltonian and 
three characteristic eigenvectors. Note the fast oscillatory behavior 
of the eigenvectors for high-energy states (insets in (c) and (d) 
panels of Fig.~\ref{fig:slika1a}). The eigenvalues of the Hamiltonian are 
indistinguishable from the analytical solution, which 
is given by \cite{Boykin}
\begin{equation}
\epsilon_k = E + 2T \cos [k \pi/(N+1)],
\label{eq:band1}
\end{equation}
where $k$ is the eigenvalue index ($k=1,\ldots,N$), that is, a cosine 
band of states of width $W \approx 4 T = 0.4 e$. The exact
solution for the periodic tight-binding chain is\cite{Ashk} 
\begin{equation}
\epsilon (k) = E + 2T \cos (2k \pi/N), \qquad (k=1,\ldots,N)
\label{eq:band2}
\end{equation}
Note the extra factor of $2$ in the argument of the cosine compared
to Eq.~(\ref{eq:band1}) and that the eigenvalue 
$\epsilon_k$ has been rewritten as $\epsilon (k)$, so that it appears as a
function of the eigenvalue index. In the periodic case, it makes sense to
characterize the eigenvalues by the wavevector, that is, $k$ becomes more 
than an eigenvalue index and has a direct interpretation in terms of the
characteristic wavelength for each eigenstate. This issue is discussed in
more detail in Appendix~\ref{sec:appendB}.

Note that Eqs.~(\ref{eq:band1}) and (\ref{eq:band2}) are not very
different for large $N$. In the limit of infinite
$N$, the band width, defined as the difference between the largest and
smallest eigenvalues, is the same in both cases, as well as the density of
states
defined as
\begin{equation}
\rho(\epsilon) = \sum_k \delta (\epsilon - \epsilon_k).
\end{equation}
There is one important difference, that is, the double 
degeneracy of states given by Eq.~(\ref{eq:band2}) for 
states with $k_1 = l$ and $k_2 = N-l$, $l=1,\ldots,N-1$, which is not the
case in Eq.~(\ref{eq:band1}).

\subsection{One defect link in a chain, simulation of scattering}
\label{sub:defect}

In this subsection, a special link 
is introduced between the 250th and 251st sites in a 
chain with total of $N=500$ sites so that $T_{250} = -0.2e$, and 
all other links are the same as before, $T_i = -0.1e$ for all $i$, 
$i \ne 250$ (the orbital energies are $E_i = e$ for all $i$, the 
same as in previous subsection). This modification of the hopping 
matrix will allow us to study the 
effects of the impurity link on the
eigenvalue spectrum and propagation of the initially localized state.

\begin{figure}[h]
\centerline{
\epsfig {file=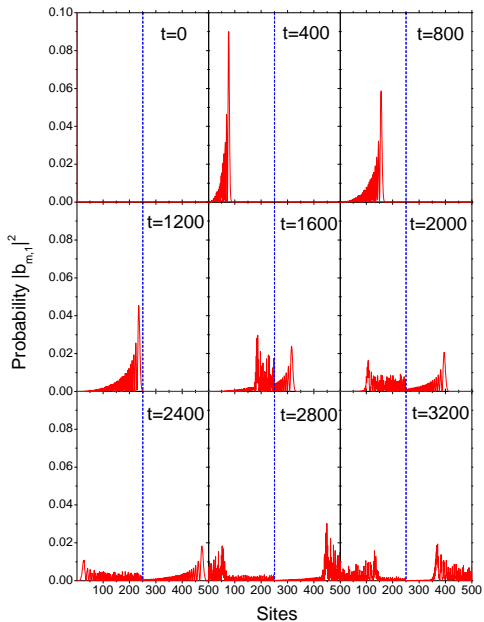,width=7cm}
}
\caption{Time dependence of the probability distribution for a 
regular chain with a special link inserted between the 250th and 
the 251st site of the chain (denoted by dashed lines in the plots). 
The time is measured in $\hbar/e$.}
\label{fig:slika2}
\end{figure}

The evolution of a state vector initially localized on the first chain site
is displayed in Fig.~\ref{fig:slika2}. Note how part of the probability
density is reflected from site 250 and 251, while the other part
continues its propagation toward the end of the chain.

\begin{figure}[h!]
\centerline{
\epsfig {file=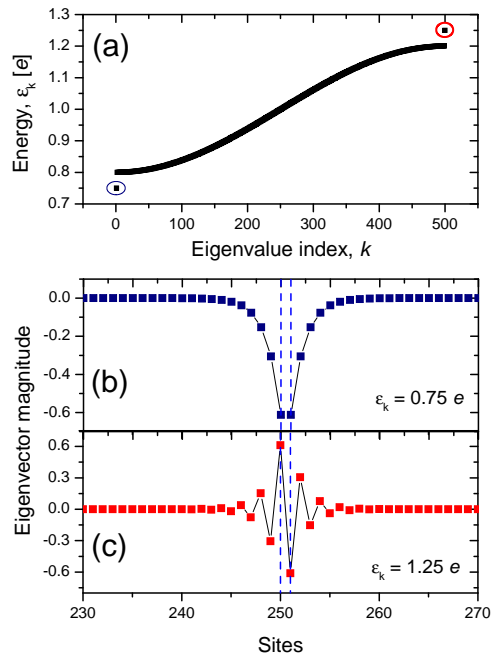,width=7cm}
}
\caption{
Eigenvalues and the two eigenvectors corresponding to impurity excitation states. 
(a) Hamiltonian eigenvalues, (b) eigenvector of 
the lowest energy eigenstate ($\epsilon_k = 0.75 e$), and (c) 
eigenvector of highest energy ($\epsilon_k = 1.25 e$). 
The two sites that are connected with an impurity link are denoted by 
dashed lines in (b) and (c).}
\label{fig:slika2a}
\end{figure}

The eigenvalue spectrum is shown in Fig.~\ref{fig:slika2a}. 
Note the appearance of two states that detach from the band. 
The eigenvectors of two special states are also displayed in 
Fig.~\ref{fig:slika2a}. The two states separated from 
the band correspond to the excitations that are localized on the 
special sites 250 and 251, that is, these states are related 
to excitations of impurity sites.

\subsection{Anderson's diagonal disorder Hamiltonian}
\label{anders}

We now study a case in which randomness is introduced in the 
Hamiltonian matrix. The orbital energies $E_i$ are given random values in a 
band of width $W$. Because these numbers are along the diagonal of the 
Hamiltonian, the model is said to have diagonal disorder. The sub-diagonal 
matrix elements are the same as before ($T_i = T = -0.1 e$). There are a 
number of interesting issues related to this model, one of which is 
called Anderson localization. The suitably modified code in Appendix~A can
be used to study the Anderson's diagonal disorder Hamiltonian. 
The number of sites is increased 
to $N=900$ and the orbital energies along the diagonal are 
given uniform random values in the interval $[e, 2e]$, that is, $W=e$.

\begin{figure}[h]
\centerline{
\epsfig {file=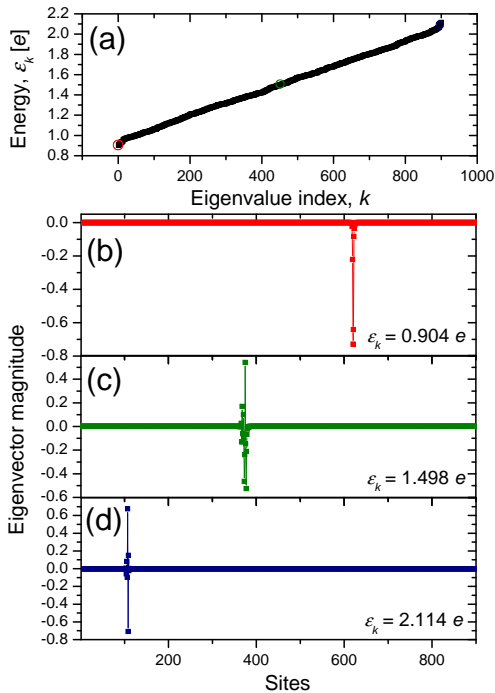,width=7cm}
}
\caption{Eigenenergies and three characteristic eigenvectors 
of the Anderson's diagonal disorder Hamiltonian.}
\label{fig:slika3}
\end{figure}

\begin{figure}[h]
\centerline{
\epsfig {file=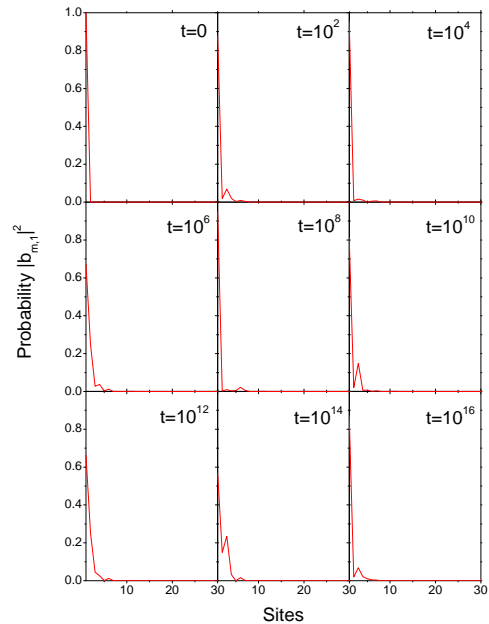,width=7cm}
}
\caption{Time dependence of the probability distribution for a 
chain with random orbital energies chosen from the interval $[e, 2e]$.
The time is measured in $\hbar/e$. Note that only the first 
30 sites of the chain with 900 sites are represented, because the site occupation probabilities are negligible beyond the 10th 
site.}
\label{fig:slika3a}
\end{figure}

One of the features of this problem is that all
the eigenstates of the Hamiltonian 
are localized.\cite{Anders1,Anders2} 
The eigenvalue spectrum and three characteristic eigenvectors 
are shown in Fig.~\ref{fig:slika3}.
The fact that all the eigenvectors are localized has a profound influence on
the propagation of an initially localized state. Figure~\ref{fig:slika3a} 
displays an evolution of a state vector initially localized on the 
first chain site. Occupation probabilities for the first 30 sites of the
chain 
are presented; the site occupation probabilities are 
negligible beyond site 10. Note that the times shown are very 
long ($t=10^{16} \hbar/e$). We conclude
that the propagation of the excitation through the 
chain is not only slow, but is effectively blocked -- the excitation
remains localized in the vicinity of the site at which it was initially
created.\cite{Anders1,Anders2} The blockage of the excitation 
propagation is related to the fact
that the projection coefficients ($a_k$ in Eq.~(\ref{eq:psi0})) of the
initially localized state have a significant magnitude only for several
eigenstates whose localization on the first site is nonvanishing. One of
those eigenvectors (the one whose maximum magnitude is on the first site)
is very similar to the initial state vector, and its projection
coefficient is the largest and close to 1. Thus, the initial state is
almost an eigenstate and its evolution is thus slow. Because
the projection coefficients $a_k$ on eigenvectors that are localized on
sites that are very distant from the first chain site are close to zero,
the propagation through the chain is essentially blocked. The total number
of the sites that become occupied during the evolution (about 5 to 10 as 
is seen in Fig.~\ref{fig:slika3a}) is related to the typical localization
width of the eigenvectors.

\subsection{Recent research on chains in the
tight-binding scheme}
\label{ssec:curr}

It is interesting to see how the
possible correlations in the diagonal disorder distribution influence
the nature of the eigenvectors and the spectrum of eigenvalues (see for
example, Refs.~\onlinecite{corrdis} and 
\onlinecite{corrdis2}). These effects could 
be studied with suitable changes of the code. The 
localization of electronic states due to disorder may drive a
metal-insulator transition (or Anderson transition) in the system, and
one-dimensional models similar to the one discussed in this article are useful in the study of random binary alloys.\cite{Naturer} In this case
there are only two characteristic orbital energies and their appearance
in the chain is random (for example, the appearance of the A orbital occurs
with probability
$p$, while the appearance of the B orbital occurs with probability
$1-p$). For a random binary alloy, the system again exhibit the
localization of eigenvectors, but correlations in the disorder may
introduce resonant states for which there is perfect electron transmission
through the system.

The electrical conduction of biological polymers, DNA in particular, has also
attracted much attention.\cite{Naturer,DNAcond} An electronic
coupling induced through the overlap of $\pi_z$ orbitals perpendicular to
the planes of the stacked base pairs in double-stranded DNA can be 
simulated using the simplified one-dimensional model in Eq.~\eqref{eq:ham},
although the realistic situation is much more complicated due to
the influence of vibrations on the distances between the base pairs
and the importance of the electronic structure of the DNA backbone. 
\cite{DNAcond,Bruinsma1,Dekker1,stretch} Models similar to the 
one studied in this article are used to study the conduction properties 
of DNA molecules.\cite{Naturer}

The model and the program in Appendix~\ref{sec:append} can be easily modified to study these
problems.
Correlations in the diagonal disorder can be studied, as well as the
introduction of disorder along the sub-diagonals (hopping matrix elements).
Localization effects can be studied in cases when the width of the diagonal
energy band $W$ is much smaller than the value of the hopping matrix
elements $T$. The opposite limit was studied in Sec.~\ref{anders}.

\appendix
\section{A program for the study of the problem}
\label{sec:append}

The following code describes the numerical solution to the problem. Due to the 
brevity
of the code, it is listed here along with comments. The source code can be
compiled as {\tt g77 -o chain1 chain1.for -llapack} 
or 
{\tt f77 -o chain1 chain1.for -llapack} \\
depending on the Fortran compiler installed. Note that the code should be
linked to the LAPACK library of routines
\cite{lapack}.
\onecolumngrid
\begin{verbatim}
1     program tba1d
2     double precision E(5000), T(5000), XI(5000,5000), 
     &       flag, work(9998), thresh, time, ReB, ImB
3     open(1, file='eigenvalues.dat', status='unknown')
4     open(2, file='eigenvectors.dat', status='unknown')
5     N = 500    
* Number of chain sites
6     thresh = 0.0D0
* Threshold for random function      
7     do i=1, N
8      E(i) = 1.0
* E - diagonal of hamiltonian matrix - orbital
9      flag = rand()
10     T(i) = -0.1
* T - subdiagonal of hamiltonian matrix - hopping
11     if (flag .lt. thresh) then
12      E(i) = 1.0 + 1.0*rand()
13     endif
14    enddo
15    call DSTEV('V',N,E,T,XI,5000,work,info)
16    do i=1, N
17     write(1,*) i, ' ',  E(i)
* On exit from DSTEV, E contains eigenvalues
18     write(2,*) 'Eigenvector: ', i
19      do j=1, N
20       write(2,*) XI(j,i)
* On exit from DSTEV, XI contains eigenvectors
21      enddo
22    enddo
23    close(1)
24    close(2)      
25    l = 1  
* Index of initially occupied site          
26    print *,'Input: time t'
27    read *, time
28    open(4, file='timedat.dat', status='unknown')
29    do j=1, N
* j counts the chain sites 
30     ReB = 0.0D0
31     ImB = 0.0D0
* Real and imaginary parts   
32     do i=1, N
* i counts the eigenstates
33      ReB = ReB + XI(l,i)*dcos(E(i)*time)*XI(j,i)
34      ImB = ImB - XI(l,i)*dsin(E(i)*time)*XI(j,i)
35     enddo            
36     write(4,*) j, ' ', ReB**2 + ImB**2
37    enddo
38    close(4)
39    end
\end{verbatim}
\twocolumngrid
The total number of sites in a chain ($N$) is defined in line 5. 
The Hamiltonian 
matrix is defined between lines 7 and 14. The parameter
{\tt thresh} allows for the introduction of random orbital 
matrix elements (lines 11--13); for {\tt thresh = 0} the orbitals 
are regular, while for {\tt thresh > 1} they are totally
random within the ranges defined by line 12 (between 1 and 2 
energy units).

The Hamiltonian matrix is diagonalized in line 15 using
the DSTEV LAPACK routine designed for the diagonalization of tridiagonal 
matrices.\cite{lapack} After the diagonalization, the $j$th column of 
the matrix {\tt XI(i,j)} contains the eigenvector corresponding to 
the $j$th eigenvalue of the Hamiltonian ($|\xi_j \rangle$). The DSTEV 
routine sorts the 
eigenvalues in an ascending order.

The time dependence of the initial state vector, 
one of the vectors from the $\{ |i \rangle \}$ basis 
in Eq.~(\ref{eq:pobasis1}), is implemented between lines 25 and 37. 
In particular, the part of the program between lines 29 and 37 
implements Eq.~(\ref{eq:tdep2}). The initial state is specified 
by the variable {\tt l} which represents the index of the 
occupied chain site at $t=0$.

The output is
written in the file {\tt timedat.dat} which can be plotted separately. This output was
used to generate Figs.~\ref{fig:slika1}, 
\ref{fig:slika2}, and \ref{fig:slika3a}. The CPU time needed to calculate
the probability distribution does not depend on the physical time input 
(variable {\tt time}) because the program does not propagate a solution in the time
domain. All the calculations needed for the evolution of the initial
state vector are performed by the diagonalization of the Hamiltonian
matrix. This output is used to calculate the probability distributions
for arbitrary times $t$ (see Eq.~(\ref{eq:tdep2})).

%%%%%%%%% THE PART BELOW WAS MOVED FROM THE MAIN TEXT
To implement the case of regular chain discused in subsection \ref{regular}, 
all randomness must be eliminated, and thus, the {\tt thresh} variable 
in line 6 is set to zero. 

To simulate one special link (impurity or defect) in the middle of 
the chain with $N=500$ sites (the case studied in subsection \ref{sub:defect}), we 
introduce the following piece of code in between the lines 13
and 14, 
\begin{verbatim}
if (i .eq. 250) then
 T(i) = -0.2
endif
\end{verbatim}
The rest of the lattice is in a regular state, that is, line 6 which controls the
amount of randomness in the chains is still 
\begin{verbatim}
thresh = 0.0D0
\end{verbatim}

For the case of totally disordered chain with $N=900$ sites studied 
in subsection \ref{anders}, lines 5 and 6 of the code should be changed to
\begin{verbatim}
5     N = 900    
6     thresh = 1.0D0
\end{verbatim}

\section{Periodic version of the regular tight-binding chain}
\label{sec:appendB}

To obtain a chain that is periodic, we have to connect the first and the
$N$th sites by the hopping matrix elements. Hence, the Hamiltonian
matrix is no longer tridiagonal because it now contains 
nonvanishing $(1,N)$ and $(N,1)$ elements. However, the Hamiltonian 
matrix is still symmetric. The periodic chain is 
easily programmed using techniques similar to those described in
Appendix~\ref{sec:append} (the code can be obtained from the author). 
For the periodic case the upper (or lower)
triangle of the Hamiltonian matrix and not only its diagonal and main
subdiagonal has to be stored.

\begin{figure}[h]
\centerline{
\epsfig {file=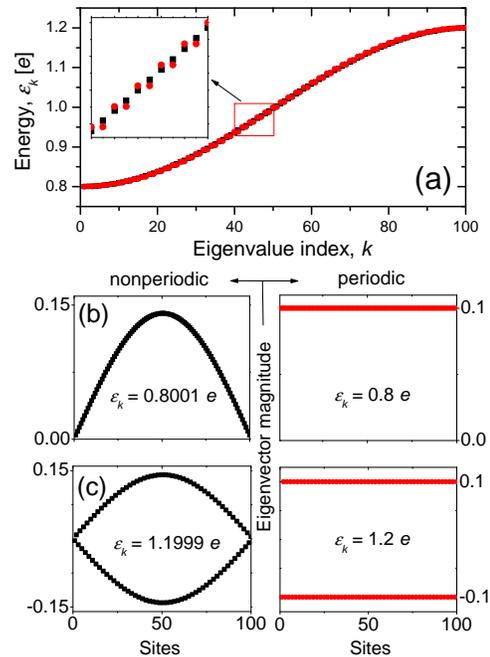,width=7cm}
}
\caption{Comparison of the eigenvalues and two characteristic 
eigenvectors for a uniform chain with non-periodic 
(squares) and periodic (circles) boundary conditions. The parameters are $E=e$, $T=-0.1e$, and $N=100$.
(a) Eigenvalues of non-periodic and 
periodic chain, 
(b) eigenvectors of 
the lowest energy eigenstates, and
(c)
eigenvectors of the highest energy eigenstates.}
\label{fig:slika7}
\end{figure}

The results for a regular periodic chain are 
shown in Fig.~\ref{fig:slika7} and compared with the 
case studied in Sec.~\ref{regular}. The parameters of 
the calculation are $E=e$, $T=-0.1e$, and $N=100$. The 
results are used to illustrate subtleties discussed at the end of 
Sec.~\ref{regular}. Note that the eigenvalue spectra look 
indistinguishable, but a closer inspection (inset in Fig.~\ref{fig:slika7}(a)) reveals that the spectrum corresponding 
to the periodic chain is doubly degenerate (except for the lowest and 
highest energy eigenvalue), which is not the case for 
the nonperiodic chain, in agreement with the discussion in
Sec.~\ref{regular}. The eigenvectors corresponding to the lowest and 
highest energy eigenstate are also different (which is true for all
eigenvectors) although the energies of these states are similar in
the two cases, see Eqs.~(\ref{eq:band1}) and 
(\ref{eq:band2})). For the periodic chain the lowest energy eigenvector
can be written in the position-occupation basis as
\begin{equation}
|\xi_{\rm lowest}\rangle = \frac{1}{\sqrt{N}} |1, 1, 1, 1 \ldots 1 \rangle,
\end{equation}
and the highest energy eigenvector in the same basis is
\begin{equation}
|\xi_{\rm highest}\rangle = \frac{1}{\sqrt{N}} |-1, 1, -1, 1 \ldots 1
\rangle.
\end{equation}


\begin{thebibliography}{}

\bibitem{goldb} A very good example is the article by A. Goldberg, 
H. M. Schey, and J. L. Schwartz, ``Computer-generated motion pictures of 
one-dimensional quantum-mechanical transmission and reflection phenomena,'' 
Am. J. Phys. {\bf 35} (3), 177--186 (1967).

\bibitem{Schiff} Leonard I. Schiff, {\em Quantum Mechanics} 
(McGraw-Hill, Singapore, 1968), 3rd ed., pp. 106--109.

\bibitem{Ashk} Neil W. Ashcroft and N. David Mermin, {\em Solid State 
Physics} (Saunders College Publishing, Fort Worth, 1976), college ed., 
pp. 176--190

\bibitem{Schiff2} Leonard I. Schiff, {\em Quantum Mechanics} 
(McGraw-Hill, Singapore, 1968), 3rd ed., pp 53.

\bibitem{Erasmo} Erasmo A. de Andrada e Silva, ``Probability current 
in the tight-binding model,'' Am. J. Phys. {\bf 60} (8), 753--754 (1992).

\bibitem{Boykin} Timothy B. Boykin and Gerhard Klimeck, ``The 
discretized Schr\"{o}dinger equation and simple models for semiconductor 
quantum wells'' Eur. J. Phys. {\bf 25} (4), 503--514 (2004).

\bibitem{Anders1} P. W. Anderson, ``Localized moments and localized 
states,'' Rev. Mod. Phys. {\bf 50} (2), 191--201 (1978).

\bibitem{Anders2} P. W. Anderson, ``Absence of diffusion in certain 
random lattices,'' Phys. Rev. {\bf 109} (5), 1492--1505 (1958).

\bibitem{corrdis} Francisco A. B. F. de Moura and Marcelo L. Lyra, 
``Delocalization in the 1D Anderson model with long-range correlated 
disorder,'' Phys. Rev. Lett. {\bf 81} (17), 3735--3738 (1998).

\bibitem{corrdis2} Hiroaki Yamada, ``Localization of electronic states 
in a nonstationary chaotic field with long-range correlation,'' 
Phys. Rev. B {\bf 69} (1), 014205-1--8 (2004).

\bibitem{Naturer} Pedro Carpena, Pedro Bernaola-Galv\'{a}n, 
Plamen Ch. Ivanov, and H. Eugene Stanley, ``Metal-insulator 
transition in chain with correlated disorder,'' Nature {\bf 418}, 
955--959 (2002).

\bibitem{DNAcond} R. G. Endres, D. L. Cox, and R. R. P. Singh, 
``The quest for high-conductance DNA,'' Rev. Mod. Phys. {\bf 76} (1), 
195--214 (2004).

\bibitem{Bruinsma1} R. Bruinsma, G. Gr\"{u}ner, M. R. D'Orsogna, and 
J. Rudnick, ``Fluctuation-facilitated charge migration along DNA,'' 
Phys. Rev. Lett. {\bf 85} (20), 4393--4396 (2000).

\bibitem{Dekker1} Gianaurelio Cuniberti, Luis Craco, Danny Porath, and 
Cees Dekker, ``Backbone-induced semiconducting behavior in short DNA 
wires,'' Phys. Rev. Lett. {\bf 65}, 241314-1--4 (2002).

\bibitem{stretch} Paul Maragakis, Ryan Lee Barnett, Efthimios Kaxiras, 
Marcus Elstner, and Thomas Frauenheim, ``Electronic structure of 
overstretched DNA,'' Phys. Rev. B {\bf 66}, 241104-1--4 (2002).

\bibitem{lapack} E. Anderson, Z. Bai, C. Bischof, J. Demmel,
J. Dongarra, J. DuCroz, A. Greenbaum, S. Hammarling, A. McKenney, 
and D. Sorensen, {\em LAPACK User's Guide} 
(SIAM Publications, Philadelphia, PA, 1999), 3rd ed., 
For more information, see \url{<www.netlib.org/lapack/>}.

\end{thebibliography}
\end{document}